\documentclass[aps,prb,reprint]{revtex4-1}

\usepackage{amsmath}
\usepackage{bm}
\usepackage{graphicx}
\usepackage{natbib}

\newcommand{\bra}[1]{\langle #1|}
\newcommand{\ket}[1]{|#1\rangle}
\newcommand{\braket}[2]{\langle #1|#2\rangle}

\begin{document}
  
\title{Entangled photons from the polariton vacuum in a switchable optical cavity}

\date{\today}

\author{Adrian Auer}
\email{adrian.auer@uni-konstanz.de}
\author{Guido Burkard}
\affiliation{Department of Physics, University of Konstanz, D-78457 Konstanz, Germany}

\begin{abstract}
  We study theoretically the entanglement of two-photon states in the ground state of the intersubband cavity system,
the so-called polariton vacuum. The system consists of a sequence of doped quantum wells located inside a microcavity
and the photons can interact with intersubband excitations inside the quantum wells. Using an explicit solution for the
ground state of the system, operated in the ultrastrong coupling regime, a post-selection is introduced, where only
certain two-photon states are considered and analyzed for mode entanglement. We find that a fast quench of the coupling
creates entangled photons and that the degree of entanglement depends on the absolute values of the in-plane wave
vectors of the photons. Maximally entangled states can be generated by choosing the appropriate modes in the
post-selection.
\end{abstract}

\maketitle

\section{\label{sec:intro}Introduction}

With the advent of quantum information theory\cite{Nielsen00}, the phenomenon of entanglement not only remained a
mysterious feature of quantum mechanics\cite{Einstein35,Schroedinger35}, but became a resource to perform tasks that are not feasible
with classical resources. Examples are quantum communication protocols, which make use of entangled
states like quantum key distribution\cite{Ekert91}, quantum teleportation\cite{Bennett93} or superdense coding\cite{Bennett92}, or the
realization of a quantum repeater\cite{Briegel98}.

Entangled photon states are often used to implement the protocols mentioned above. Today, there exist several different
proposals for the production of bipartite entangled photon states, most prominently type-II parametric
down-conversion\cite{Kwiat95} and biexciton decay in a quantum dot\cite{Benson00}.

The fundamental requirements for such a photon-pair source to be used in quantum information processing are that
the states have to possess a sufficient amount of entanglement and that the production of the two photons has to be
deterministic and efficient. Determinism means that the release of the photons can be triggered by some external control
parameter. Efficiency means that the probability for this event is near unity.

Here, we study the intersubband cavity system, for which the emission of correlated photon pairs was predicted
theoretically\cite{Ciuti05} and can be triggered by modulating the light-matter interaction between microcavity
photons and electronic excitations in the quantum wells. Those intersubband transitions are mainly used in quantum
well infrared photodetectors\cite{Liu99a} and quantum cascade lasers\cite{Liu99b,Faist1994,Koehler02,Colombelli03}.
Embedded in a microcavity, it is possible to reach a regime of ultrastrong light-matter
coupling\cite{Gunter09,Anappara09,Todorov10,Geiser11}, in which
the vacuum-field Rabi frequency can be of the order of the intersubband transition frequency and the ground state of
the system, a squeezed vacuum, contains already a non-zero number of photons. Another type of system, which can
reach the ultrastrong coupling regime as well, are superconducting circuits\cite{Bourassa09,Niemczyk10,Forn10},
where the emission of quantum vacuum radiation was just recently demonstrated \cite{Wilson11}.

In this paper, we analyze the ground state of the intersubband cavity system, the so-called polariton vacuum,
related to two-photon entanglement. We use an explicit expression for the polariton vacuum and, after
post-selecting certain photonic states, quantify the mode entanglement between the photon pairs via the concurrence.

\section{\label{sec:sys}The intersubband cavity system}
The intersubband cavity system was studied theoretically by Ciuti \textit{et al.} \cite{Ciuti05}.
It consists of $n_\text{QW}$ identical quantum wells embedded inside a semiconductor optical microcavity. 
The quantum wells are assumed to be negatively charged with a two-dimensional electron gas (2DEG)
with density $N_\text{2DEG}$ that populates the first subband. We consider the interaction
of intersubband excitations between the two lowest subbands and photons of the fundamental cavity
mode.
\begin{figure}
  \includegraphics[width=\columnwidth]{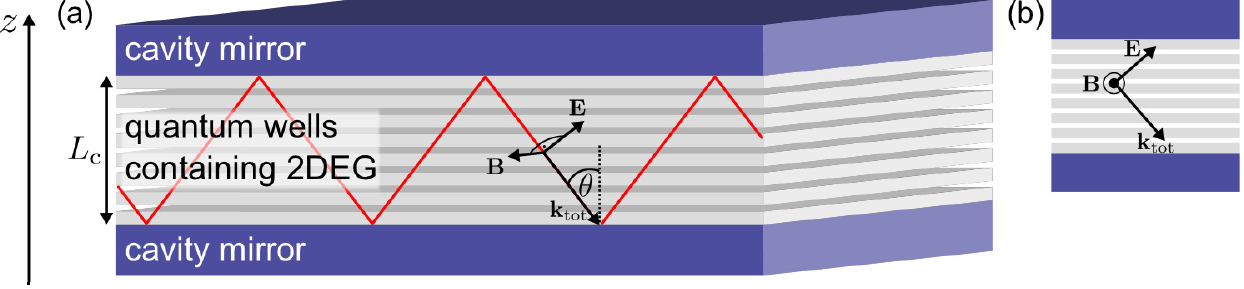}
  \caption{\label{fig:system}(a) The system under consideration: A sequence of doped QWs (gray) inside a microcavity of
  length $L_\text{c}$ (cavity mirrors in blue) form the intersubband cavity system. The light-matter interaction in such
  a system depends on the propagation angle $\theta$ of the cavity photons. The transverse-magnetic polarization is
  indicated by the electric and magnetic field vectors. The electric field $\bm E$ lies in the plane of incidence and
  the magnetic field $\bm B$ is perpendicular to it, and both are perpendicular to the photon wave vector ${\bm
  k}_\text{tot}$. (b) The plane of incidence to better demonstrate the TM-polarization of the photons.}
\end{figure}
\begin{figure}
  \includegraphics[width=\columnwidth]{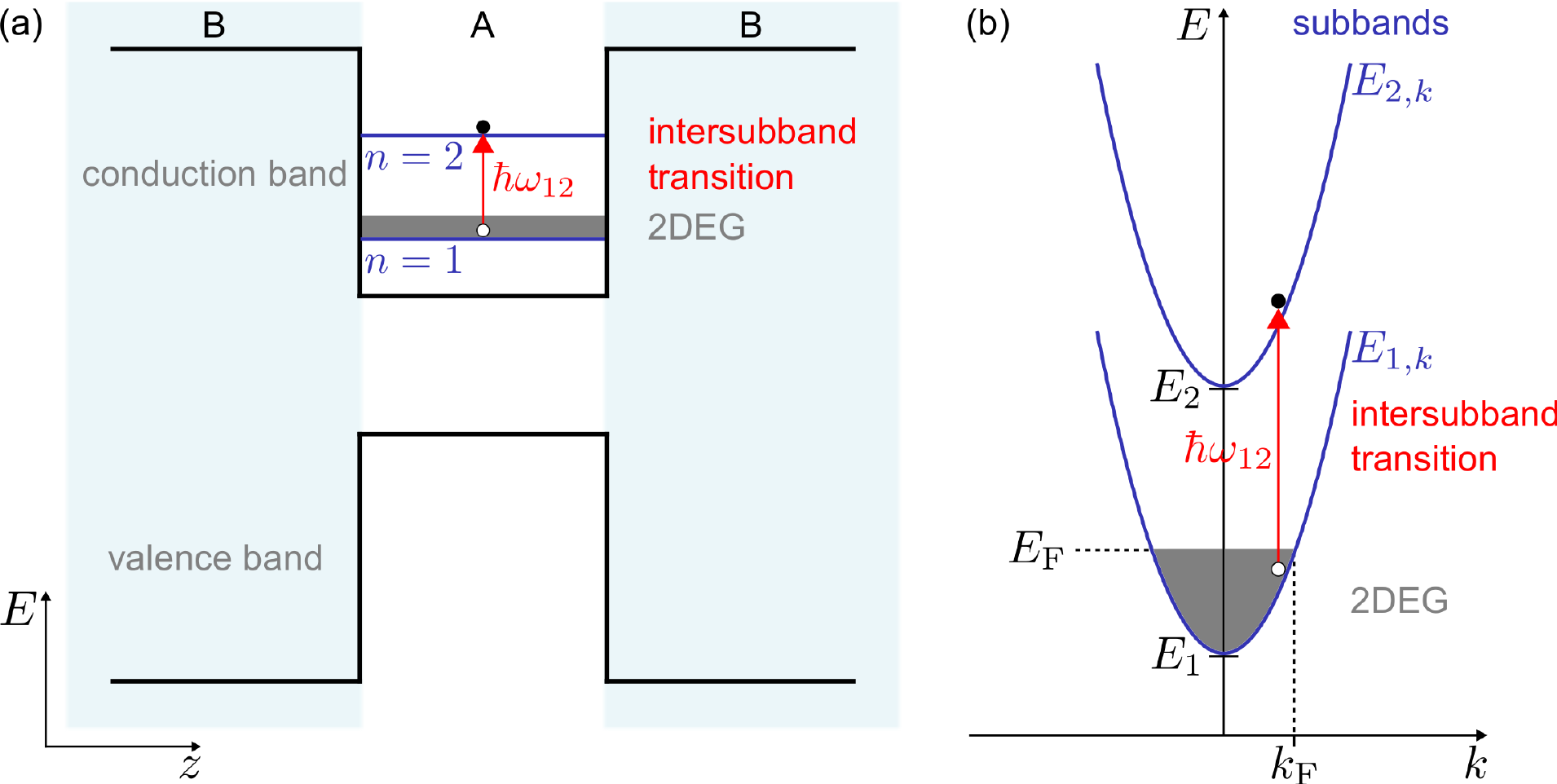}
  \caption{\label{fig:trans_sub}Subband energy structure of a quantum well (QW) formed in a semiconductor
  heterostructure. (a) In real space, along the growth direction $z$ of the structure, the semiconductors forming
  the QW are denoted as A and B. The QW contains
  a two-dimensional electron gas (2DEG). Here, we study intersubband transitions between the first two subbands $n = 1$ and
  $n = 2$ with transition energy $\hbar \omega_{12}$. (b) The same situation in $k$ space, where the 2DEG populates states up to the Fermi energy $E_\text{F}$,
  with a corresponding wave vector $k_\text{F}$.}
\end{figure}

The ultrastrong coupling regime, in which the vacuum-field Rabi splitting is of the order of
the intersubband transition energy, can be reached due to the large dipole moment\cite{Gunter09} of intersubband
transitions and the collective coupling to all electrons of the 2DEGs.
In this regime, the rotating wave approximation \cite{Vogel06} is not valid anymore and the full
light-matter interaction Hamiltonian including the antiresonant terms has the form \cite{Ciuti05}
\begin{equation}
\label{eqn:hamiltonian}
  H = H_0 + H_\text{res} + H_\text{anti},
\end{equation}
which consists of the three terms
\begin{eqnarray}
  H_0 &=& \sum \limits_{\bm k} \hbar \omega_\text{c}(k) \left( a_{\bm k}^\dagger a_{\bm k}^{\phantom{\dagger}}
  + \frac{1}{2} \right) + \sum \limits_{\bm k} \hbar \omega_{12} b_{\bm k}^\dagger b_{\bm k}^{\phantom{\dagger}}, \\
  H_\text{res} &=& \sum \limits_{\bm k} i \hbar \Omega_\text{R}(k) \left( a_{\bm k}^\dagger 
  b_{\bm k}^{\phantom{\dagger}} - a_{\bm k}^{\phantom{\dagger}} b_{\bm k}^\dagger \right) \nonumber \\
  &+& \sum \limits_{\bm k} \hbar D(k) \left( a_{\bm k}^\dagger 
  a_{\bm k}^{\phantom{\dagger}} + a_{\bm k}^{\phantom{\dagger}} a_{\bm k}^\dagger \right),\\
  \label{eqn:anti}
  H_\text{anti} &=& \sum \limits_{\bm k} i \hbar \Omega_\text{R}(k) \left( a_{\bm k}^{\phantom{\dagger}}
  b_{-\bm k}^{\phantom{\dagger}} - a_{\bm k}^\dagger b_{- \bm k}^\dagger \right) \nonumber \\
  &+& \sum \limits_{\bm k} \hbar D(k) \left( a_{\bm k}^{\phantom{\dagger}} 
  a_{-\bm k}^{\phantom{\dagger}} + a_{\bm k}^\dagger a_{-\bm k}^\dagger \right),
\end{eqnarray}
where $H_0$ describes the uncoupled photonic and electronic systems, $H_\text{res}$ is the resonant part of
the light-matter interaction and the antiresonant terms, usually neglected under the rotating wave approximation,
are given by $H_\text{anti}$.

The operator $a_{\bm k}^{(\dagger)}$ annihiltes (creates) a cavity photon with in-plane wave vector
${\bm k} = (k_x, k_y)$ and transverse-magnetic (TM) polarization. The reason why only this photon polarization is considered,
is the selection rule for intersubband transitions \cite{Liu99a}. The dipole moment points along the growth direction,
so the exciting radiation must have a finite electric field component in $z$-direction. As can be seen from 
Fig.~\ref{fig:system}, TM-polarized light where the magnetic field component is perpendicular to the plane of incidence,
has an electric field component in $z$-direction when it propagates under a finite angle $\theta$. However, if the
polarization is not purely transverse-magnetic, it is still sufficient to consider only one polarization direction,
since only the TM-polarized part of the radiation is absorbed. 

The only bright collective intersubband excitation with in-plane wave vector $\bm k$ is described by the
operators $b_{\bm k}^{(\dagger)}$ that fulfill Bose commutation relations $[b_{\bm k},b_{{\bm k}'}^\dagger] \simeq
\delta_{{\bm k},{\bm k}'}$ in the weak excitation regime, i.e. when the number of intersubband excitations
is much less than the number of electrons forming the 2DEG \cite{Ciuti05,DeLiberato09}.

The dispersion of the fundamental cavity mode $\omega_\text{c}(k)$ is given by
\begin{equation}
  \omega_\text{c}(k) = \frac{c}{\sqrt{\varepsilon}}\sqrt{k^2 + k_z^2},
\end{equation}
where c is the speed of light, $\varepsilon$ is the dielectric constant of the material used as
cavity spacer and the quantization of $k_z$ can depend in a complicated way on the boundary conditions. In the following,
$k = |{\bm k}|$ is the length of the in-plane wave vector.

The vacuum Rabi frequency $\Omega_\text{R}(k)$ for the intersubband cavity system is given by
\cite{Ciuti05,Hagenmueller10}
\begin{equation}
  \Omega_\text{R}(k) = \left( \frac{e^2 N_\text{2DEG} n_\text{QW}^\text{eff} f_{12} \omega_{12}}{2 \varepsilon_0
\varepsilon m_0 L_\text{c}^\text{eff} \omega_\text{c}(k)}
   \sin^2 \theta(k) \right)^\frac{1}{2}.
\end{equation}
Here $e$ is the elementary charge, $\varepsilon_0$ is the vacuum permittivity and $m_0$ is the free electron
mass. $L_\text{c}^\text{eff}$ denotes an effective cavity thickness that depends on the type of cavity mirrors and
$n_\text{QW}^\text{eff}$ is an effective number of embedded quantum wells since not all quantum wells are equally
coupled to the photon field. $f_{12}$ is the oscillator strength of the transition between the subbands.
For a deep rectangular well, one can use the approximation $f_{12} \simeq m_0 / m^*$ ($m^*$: effective electron
mass) \cite{Liu99a,Ciuti05}. Finally, $\theta(k)$ is the propagation angle of a cavity photon: $\sin \theta(k) =
k/\sqrt{k^2 + k_z^2}$. The dispersive coupling parameter $D(k)$ can be approximated by $D(k) \simeq \Omega_\text{R}^2(k) / 
\omega_{12}$, which is valid for deep rectangular wells and exact for parabolic well potentials \cite{Ciuti05}.

\begin{figure}
  \includegraphics[width=\columnwidth]{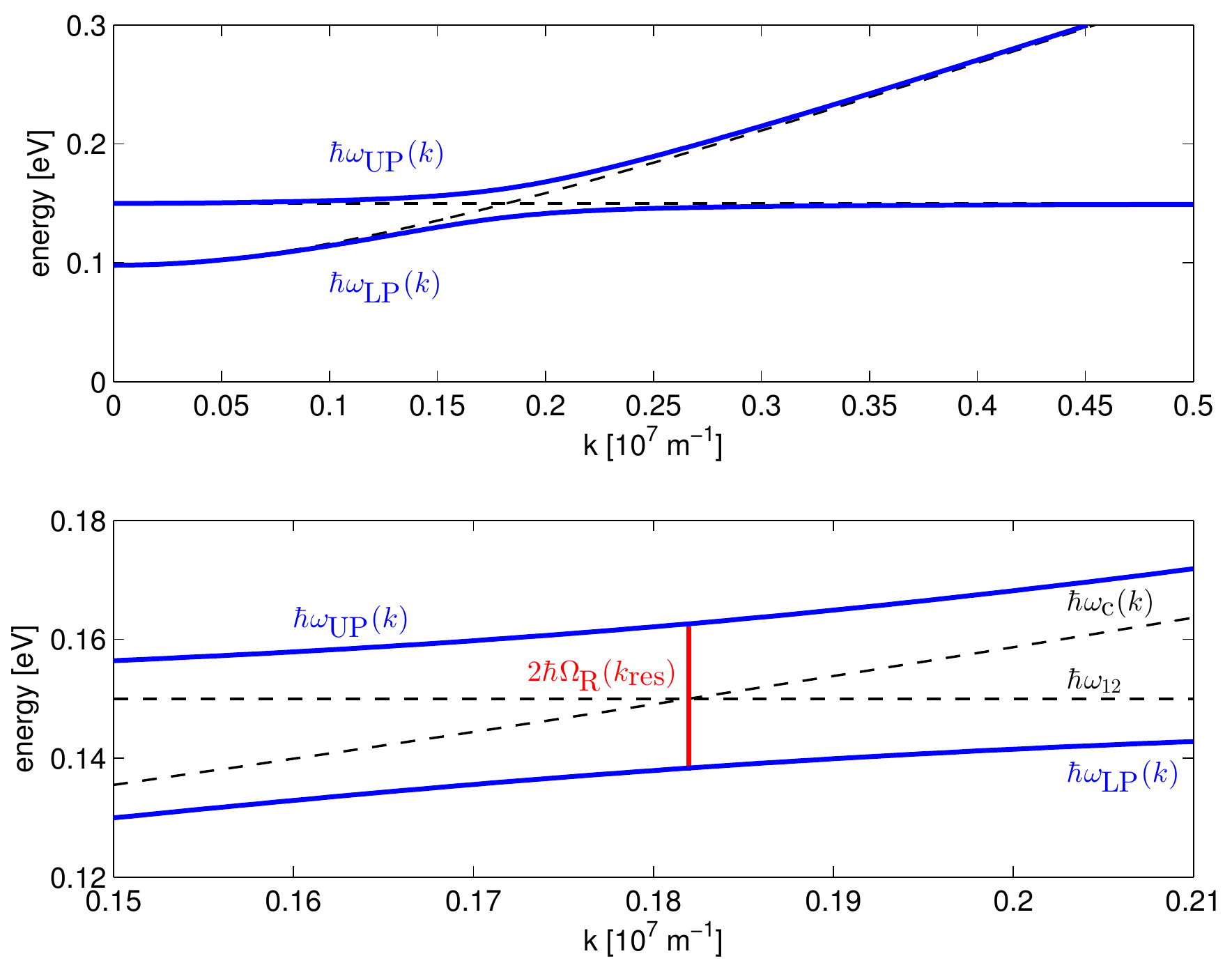}
  \caption{\label{fig:dispersion}The polariton energy dispersions $\hbar \omega_\text{LP}(k)$ and $\hbar
  \omega_\text{UP}(k)$ as a function of the absolute value of the in-plane wave vactor $k$. Here, $n_\text{QW}^\text{eff} = 50$
  GaAs/AlGaAs ($\varepsilon = 10.0$, $f_{12} = 14.9$) QWs are assumed to be located inside a cavity of length
  $L_\text{c}^\text{eff} = 2\,\mu\text{m}$ and doped with a two-dimensional electron gas density of $N_\text{2DEG} =
  10^{12}\, \text{cm}^{-2}$. The lower plot shows the anticrossing region where the dotted lines are the bare cavity
  photon dispersion $\hbar \omega_\text{c}(k)$ and the constant intersubband transition energy $\hbar \omega_{12} =
  150\,\text{meV}$. The vacuum-field Rabi splitting is $2 \hbar \Omega_\text{R}(k_\text{res})$.}
\end{figure}

The Hamiltonian (\ref{eqn:hamiltonian}) can be diagonalized with an extended Bogoliubov transformation \cite{Hopfield58},
also known as Hopfield transformation \cite{Quattropani86}, where new bosonic operators
\begin{equation}
\label{eqn:Bogoliubov_trans}
  p_{j,{\bm k}} = w_j(k)a_{\bm k} + x_j(k)b_{\bm k} + y_j(k)a_{-\bm k}^\dagger
  + z_j(k)b_{-\bm k}^\dagger
\end{equation}
are introduced that describe a quasiparticle called intersubband cavity polariton\cite{Dini03} and $j$
indicates wether it belongs to the lower ($j = \text{LP}$) or upper ($j = \text{UP}$) polariton branch.
The wave vectors $\bm k$ are still meant to be in-plane. By an appropriate choice
of the Hopfield coefficients $w_j(k)$, $x_j(k)$, $y_j(k)$ and $z_j(k)$, which are already taken to depend only on $k$, the
Hamiltonian becomes diagonal
\begin{equation}
  H = E_\text{G} + \sum \limits_{j \in \{\text{LP,UP}\}} \sum \limits_{\bm k} \hbar \omega_j(k)
  p_{j,{\bm k}}^\dagger p_{j,{\bm k}}.
\end{equation}
Here, $E_\text{G}$ denotes the ground state energy. The resulting lower and and upper polariton dispersion are shown in
Fig.~\ref{fig:dispersion}. Here, as well as for all further quantitative
resultes, we assume GaAs/AlGaAs quantum wells, which have been commonly used experimentally \cite{Dini03,Anappara05,Anappara06,Gunter09}.
Hence, the material-dependent parameters are $f_{12} = 14.9$ ($m_\text{GaAs}^* = 0.067$) and $\varepsilon = 10$. Furthermore,
the number of embedded quantum wells $n_\text{QW}$, the length of the microcavity $L_\text{c}$ and the intersubband
transition frequency $\omega_{12}$, which is determined by the quantum well depth and thickness, can be adjusted
during the manufacturing process. The density of the two-dimensional electron gas can be varied experimentally. To obtain
the results of Fig.~\ref{fig:exp_coeff}, we chose $n_\text{QW}^\text{eff} = 50$, $L_\text{c}^\text{eff} = 2
\,\mu\text{m}$, $\hbar
\omega_{12} = 150\,\text{meV}$
 and $N_\text{2DEG} = 10^{12}\, \text{cm}^{-2}$ as
one particular set of experimentally reasonable values of the parameters mentioned above.

The ground state of the intersubband cavity system, operating in the ultrastrong coupling regime, is not
the ordinary vacuum $\ket{0}$, in which there are no cavity photons and no intersubband excitations
present
\begin{equation}
  a_{\bm k} \ket{0} = b_{\bm k} \ket{0} = 0,
\end{equation}
but a state $\ket{G}$ that exhibits no intersubband cavity polaritons
\begin{equation}
\label{eqn:def_vac}
  p_{j,{\bm k}} \ket{G} = 0,\quad j \in \{\text{LP,UP}\}.
\end{equation}
Without knowing the explicit form of $\ket{G}$, one can show that the ground state has some peculiar properties
in the ultrastrong coupling regime that were worked out in Ref.~\onlinecite{Ciuti05}, whereof the essential ones
are that it contains a finite number of photons:
\begin{equation}
  \bra{G} a_{\bm k}^\dagger a_{\bm k} \ket{G} = \left| y_\text{LP}(k) \right|^2 + \left| y_\text{UP}(k) \right|^2,
\end{equation}
and photons with opposite in-plane wave vectors $\bm k$ and $-\bm k$ are correlated:
\begin{equation}
\label{eqn:correlations}
  \bra{G} a_{\bm k} a_{-\bm k} \ket{G} = - w_\text{LP}^*(k) y_\text{LP}(k) - w_\text{UP}^*(k) y_\text{UP}(k).
\end{equation}
One can see that only if the light-matter interaction is so strong that the Hopfield coefficients $y_\text{LP}(k)$
and $y_\text{UP}(k)$ are reasonably large, $| y_j(k) |^2 \sim 0.1$ (i.e. when the antiresonant terms 
(\ref{eqn:anti}) cannot be neglected and therefore the extended Bogoliubov transformation (\ref{eqn:Bogoliubov_trans})
is necessary), the ground state $\ket{G}$ differs significantly from the vacuum state $\ket{0}$.

The idea is now that the correlations (\ref{eqn:correlations}) can lead to entanglement of two photons propagating
in opposite directions. These photons are, however, virtual excitations, but it is conjectured \cite{Ciuti05,Ciuti06,
DeLiberato07} that they can be released by a non-adiabatic switch-off (quench) of the vacuum Rabi frequency
$\Omega_\text{R}(k)$. An experimental approach to this scenario is an ultrafast change
of the density $N_\text{2DEG}$ of two-dimensional electron gas\cite{Gunter09,Anappara05,Anappara06}. One mechanism to achieve a modulation
of the parameter $N_\text{2DEG}$ is a gate voltage, which can lead to the depletion of the QWs \cite{Anappara05}. The
rapidity is restricted by capacitance of the gates, however. Another implementation uses two asymmetrically coupled QWs,
in which one QW can be charged by electron tunneling and this process can happen on the picosecond time scale or
faster\cite{Anappara06}. A promising idea to achieve an ultrafast coupling modulation is an all-optical
control scheme\cite{Gunter09}, in which electrons from the valence band are resonantly excited to the first subband by a
femtosecond laser pulse. In this manner, it could be demonstrated experimentally
that the coupling between the cavity photon field and the intersubband transitions in the quantum wells can be switched on
in a time shorter than a cycle of light in the microcavity. In Ref.~\onlinecite{Ciuti06,DeLiberato07}
the spectrum of the radiation exiting the cavity was derived in more detailed calculations, when a time-dependent coupling
$\Omega_\text{R}(k,t)$ is predominant in the system and it is predicted that the vacuum radiation rises above the black body
radiation.

\section{\label{sec:sol}Exact ground state}

A pioneering calculation of the polariton ground state of a bulk dielectric was given by Quattropani \textit{et al.}\cite{Quattropani86}.
The solution is given by independent photon and polarization states. Since the Hamiltonian of the intersubband cavity system
is similar to the one in Ref.~\onlinecite{Quattropani86}, we use their treatment to determine the explicit form of the
polariton
vacuum $\ket{G}$ being the ground state of the Hamiltonian (\ref{eqn:hamiltonian}) of the intersubband cavity system. The
difference is just that the sums in (\ref{eqn:hamiltonian}) cover all in-plane wave vectors and the intersubband transition
frequency $\omega_{12}$ is taken to be dispersionless.

The ansatz for the polariton vacuum $\ket{G}$ is:
\begin{equation}
  \ket{G} = \frac{1}{N} e^{\frac{1}{2} \sum \limits_{\bm k} \left[ G(k) \left( a_{\bm k}^\dagger a_{-\bm k}^\dagger +
  b_{\bm k}^\dagger b_{-\bm k}^\dagger \right) + F(k) \left( a_{\bm k}^\dagger b_{-\bm k}^\dagger +
  b_{\bm k}^\dagger a_{-\bm k}^\dagger \right)\right] } \ket{0}.
\end{equation}
$N$ is a normalization constant and the expansion coefficients $G(k)$ and $F(k)$ have to be determined in order to
satisfy the definition of the polariton vacuum (\ref{eqn:def_vac}):
\[
  p_{j,{\bm k}} \ket{G} = 0,\quad j \in \{\text{LP,UP}\}.
\]
We anticipate that the functions $G(k)$ and $F(k)$ will only depend on the absolute value of the in-plane wave vector
$\bm k$. After some algebra using commutation relations, which is explicitly given in Ref.~\onlinecite{Quattropani86},
the action of $a_{\bm k}$ and $b_{\bm k}$ on $\ket{G}$ is, e.g.:
\begin{eqnarray}
\label{eqn:act_a}
  a_{\bm k} \ket{G} &=& \left( G(k) a_{-\bm k}^\dagger + F(k) b_{-\bm k}^\dagger \right) \ket{G}, \\
\label{eqn:act_b}
  b_{\bm k} \ket{G} &=& \left( G(k) b_{-\bm k}^\dagger + F(k) a_{-\bm k}^\dagger\right) \ket{G}.
\end{eqnarray}
Inserting (\ref{eqn:act_a}) and (\ref{eqn:act_b}) into the definitions (\ref{eqn:Bogoliubov_trans}) and (\ref{eqn:def_vac}),
one obtains a system of equations for the coefficients $G(k)$ and $F(k)$:
\begin{eqnarray}
  w_j(k) G(k) + x_j(k) F(k) + y_j(k) &=& 0, \\
  w_j(k) F(k) + x_j(k) G(k) + z_j(k) &=& 0,
\end{eqnarray}
which has the solutions:
\begin{eqnarray}
  G(k) &=& \frac{x_j(k) z_j(k) - w_j(k) y_j(k)}{w_j^2(k) - x_j^2(k)}, \\
  F(k) &=& \frac{x_j(k) y_j(k) - w_j(k) z_j(k)}{w_j^2(k) - x_j^2(k)}.
\end{eqnarray}
That this can be fulfilled simultaneously by the Hopfield coefficients of the lower and upper polariton, can
be seen from the following relations \cite{Hopfield58,Quattropani86}:
\begin{eqnarray}
  w_\text{LP}(k) = x_\text{UP}(k)&,& \quad x_\text{LP}(k) = w_\text{UP}(k), \nonumber \\
  y_\text{LP}(k) = z_\text{UP}(k)&,& \quad z_\text{LP}(k) = y_\text{UP}(k).
\end{eqnarray}
By inserting the explicit expressions of the Hopfield coefficients, the expansion coefficients can be rewritten
as \cite{Quattropani86}:
\begin{eqnarray}
\label{eqn:exp_coeff}
  G(k) &=& \frac{\omega_{12} + \omega_\text{c}(k) - \omega_\text{LP}(k) - \omega_\text{UP}(k)}{\omega_{12} -
  \omega_\text{c}(k) - \omega_\text{LP}(k) - \omega_\text{UP}(k)}, \\
\label{eqn:exp_coeff_F}
  F(k) &=& -i\frac{\omega_{12}}{\Omega_\text{R}(k)} G(k).
\end{eqnarray}
Finally, the polariton vacuum $\ket{G}$ is calculated to be
\begin{equation}
  \ket{G} = \frac{1}{N} e^{\frac{1}{2} \sum \limits_{\bm k} G(k) \left( a_{\bm k}^\dagger a_{-\bm k}^\dagger +
  b_{\bm k}^\dagger b_{-\bm k}^\dagger - 2 i \frac{\omega_{12}}{\Omega_\text{R}(k)} a_{\bm k}^\dagger b_{-\bm k}^\dagger
  \right)} \ket{0},
\end{equation}
because the last two terms in the exponential can be combined, and with the normalization $N$ given by
\begin{equation}
  N = \prod \limits_{\bm k} \left( \left|w_\text{LP}(k)\right|^2 + \left|x_\text{LP}(k)\right|^2 \right)^\frac{1}{2}.
\end{equation}

\begin{figure}
  \includegraphics[width=\columnwidth]{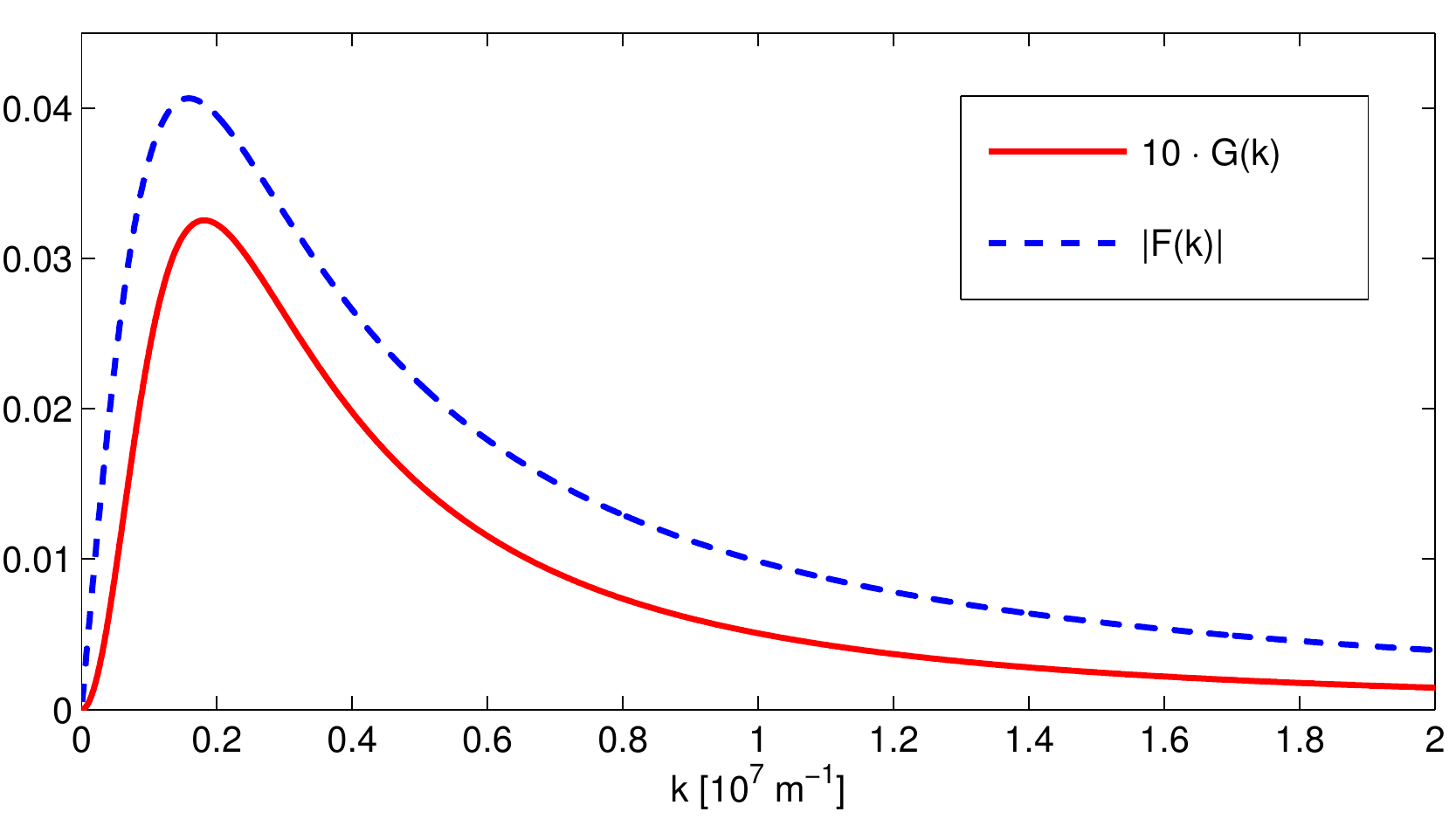}
  \caption{\label{fig:exp_coeff}The absolute values of the expansion coefficients $G(k)$ and $|F(k)|$ as a function of
  the length $k$ of the in-plane wave vector. (Parameter values: $\varepsilon = 10$, $f_{12} = 14.9$,
  $n_\text{QW}^\text{eff} = 50$, $L_\text{c}^\text{eff} = 2\,\mu\text{m}$, $\hbar \omega_{12} = 150\,\text{meV}$,
  $N_\text{2DEG} = 10^{12}\,\text{cm}^{-2}$.)}
\end{figure}

The dependance of $G(k)$ and $|F(k)|$ on $k$ is plotted in Fig.~\ref{fig:exp_coeff}.
One can see that $|F(k)|$ is about one order of magnitude larger than $G(k)$ and
\begin{equation}
  G(k) \approx |F(k)|^2 \ll 1,
\end{equation}
in which this was checked for a wide range of experimentally acceptable values of the parameters $n_\text{QW}$, $L_\text{c}$,
$\omega_{12}$ and $N_\text{2DEG}$. Therefore, the polariton vacuum state $\ket{G}$ will be expanded in a Taylor
series to the second order in these coefficients as an approximation.

If the light-matter interaction is turned off ($\Omega_\text{R}(k) = 0$), then $G(k)$ and $F(k)$ are zero, so the ground
state would be the ordinary vacuum $\ket{0}$, as expected.

\section{\label{sec:ent}Photon entanglement}

As seen in the previous section, the intersubband cavity system contains a finite number of photons if it is in the
ultrastrong coupling regime. One possibility of a photon pair in the ground state $\ket{G}$ to be entangled will be
studied below. After specifying the type of entanglement, it is quantified by the concurrence
\cite{Hill97,Wootters98}.

\subsection{\label{subsec:idea}Mode entanglement}

We first define the type of photon entanglement. Since the transverse-magnetic polarization of the interacting photons
is fixed by the selection rule for intersubband transitions \cite{Liu99a}, polarization entanglement as achieved with
parametric down-conversion or the biexciton decay is out of the question. But there exist anomalous correlations between
photons with opposite in-plane momentum (\ref{eqn:correlations})
\[
  \bra{G} a_{\bm k} a_{-\bm k} \ket{G} = - w_\text{LP}^*(k) y_\text{LP}(k) - w_\text{UP}^*(k) y_\text{UP}(k).
\]
Our idea is to test the photonic states in $\ket{G}$ for mode or frequency entanglement \cite{Bouwmeester01}.

\begin{figure}
  \includegraphics[width=\columnwidth]{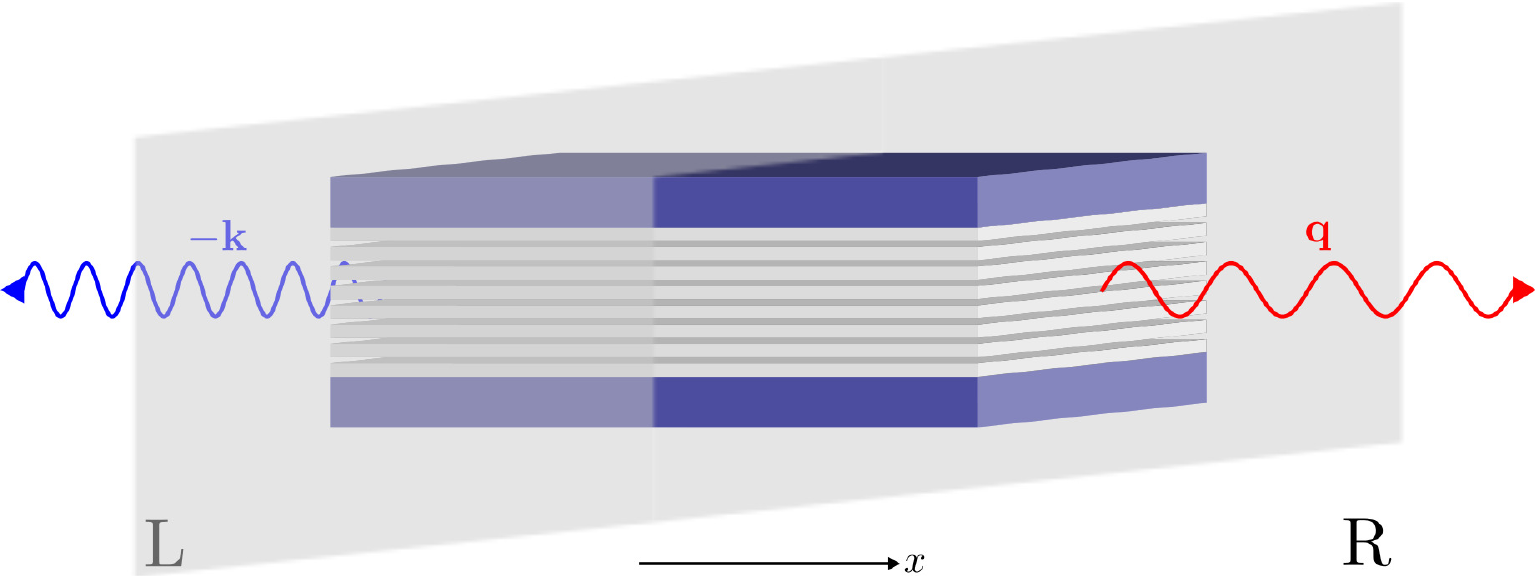}
  \caption{\label{fig:entangle}Two photons with different frequencies (colors) leaving the cavity in
  opposite directions. The two subsystems left (L) and right (R) are defined via the sign of $k_x$. Since there is a difference in
  frequency, the photons have different in-plane wave vectors.}
\end{figure}

In the following, we limit the treatment to only one direction, since the correlations (\ref{eqn:correlations})
occur only for photons with exactly opposite in-plane wave vectors. This direction is chosen to be the
$x$-direction. This is shown schematically in Fig.~\ref{fig:entangle}. Photons with a negative (positive) $x$-component
of the wave vector belong to subsystem L(R) for left(right).

$\ket{G}$ itself is a vector from a Hilbert space $\mathcal{H}$, which has a tensor product structure $\mathcal{H} = \mathcal{F}^a
\otimes \mathcal{F}^b$, where $\mathcal{F}^a$ and $\mathcal{F}^b$ denote the Fock spaces of the photons and intersubband excitations,
respectively. The situation depicted in Fig.~\ref{fig:entangle} is described by vectors in a subspace
$\mathcal{H}_\text{LR} \subset \mathcal{ H}$, which is itself a tensor product of $\mathcal{H}_\text{L}$ and
$\mathcal{H}_\text{R}$,
the Hilbert spaces of the subsystems L and R, $\mathcal{H}_\text{LR} = \mathcal{H}_\text{L} \otimes \mathcal{H}_\text{R}$.
Since $\ket{G}$ contains states from outside $\mathcal{H}_\text{LR}$, we project onto $\mathcal{H}_\text{LR}$
with an appropriate projection operator, which will be described in the following and  could be realized
experimentally by a post-selective measurement.

\subsection{\label{subsec:post}Post-selection}

The post-selection needs to fulfill the following requirements: first, we only allow for states in which two photons with
opposite in-plane wave vectors appear. In addition, the post-selection is even more restrictive in terms of
allowed modes. We only consider two different modes in each subsystem L and R, respectively, that have, however
the same absolute value of the in-plane wave vector. I.e. we choose $k$ and $q$, with $k \neq q$, and consider the modes $\bm k = (k,0)$
and $-\bm k = (-k,0)$ and accordingly $\bm q = (q,0)$ and $-\bm q = (-q,0)$, where all the wave vectors point along the
$x$-direction. So the basis states of $\mathcal{H}_\text{L}$ are
\begin{eqnarray}
  \ket{\bm k}_\text{L} &=& a_{-\bm k}^\dagger \ket{0}_a, \\
  \ket{\bm q}_\text{L} &=& a_{-\bm q}^\dagger \ket{0}_a,
\end{eqnarray}
where $\ket{0}_a$ is the photon vacuum. The basis of $\mathcal{H}_\text{R}$ is
\begin{eqnarray}
  \ket{\bm k}_\text{R} &=& a_{\bm k}^\dagger \ket{0}_a, \\
  \ket{\bm q}_\text{R} &=& a_{\bm q}^\dagger \ket{0}_a.
\end{eqnarray}
Hence, one possible product basis of $\mathcal{H}_\text{LR}$ is
\begin{eqnarray}
\label{eqn:basis}
  \ket{\bm k}_\text{L} \otimes \ket{\bm k}_\text{R}&,& \quad \ket{\bm k}_\text{L} \otimes \ket{\bm q}_\text{R}, \nonumber \\
  \ket{\bm q}_\text{L} \otimes \ket{\bm k}_\text{R}&,& \quad \ket{\bm q}_\text{L} \otimes \ket{\bm q}_\text{R}.
\end{eqnarray}

For all further calculations, the polariton vacuum $\ket{G}$ is expanded to the second order in the small
expansion coefficient $G(k)$,
  \begin{eqnarray}
    \ket{G} &=& \frac{1}{N} e^{\frac{1}{2} \sum \limits_{\bm k} G(k) \left( a_{\bm k}^\dagger a_{-\bm k}^\dagger +
  b_{\bm k}^\dagger b_{-\bm k}^\dagger - 2 i \frac{\omega_{12}}{\Omega_\text{R}(k)} a_{\bm k}^\dagger b_{-\bm k}^\dagger \right)}
  \ket{0} \nonumber \\
  &\approx& \frac{1}{\widetilde{N}}\left[ 1 + \frac{1}{2} \sum \limits_{\bm k} G(k) \mathcal{T}_{\bm k}^\dagger
  + \frac{1}{8} \left( \sum \limits_{\bm k} G(k) \mathcal{T}_{\bm k}^\dagger \right)^2 \right] \ket{0} \nonumber \\
  &\equiv& \ket{G^{(2)}},
  \end{eqnarray}
where $\widetilde{N}$ is a new normalization constant to preserve $\braket{G^{(2)}}{G^{(2)}} = 1$ and the operator
$\mathcal{T}_{\bm k}^\dagger$ is
\begin{eqnarray}
  \mathcal{T}_{\bm k}^\dagger = a_{\bm k}^\dagger a_{-\bm k}^\dagger +
  b_{\bm k}^\dagger b_{-\bm k}^\dagger - 2 i \frac{\omega_{12}}{\Omega_\text{R}(k)} a_{\bm k}^\dagger b_{-\bm k}^\dagger.
\end{eqnarray}
Thus, the two-photon states with opposite in-plane wave vector, i.e. the states that fulfill the post-selection
requirements, in linear order are
\begin{eqnarray}
  a_{\bm k}^\dagger a_{-\bm k}^\dagger \ket{0}_a &=& \ket{\bm k}_\text{L} \ket{\bm k}_\text{R}, \\
  a_{\bm q}^\dagger a_{-\bm q}^\dagger \ket{0}_a &=& \ket{\bm q}_\text{L} \ket{\bm q}_\text{R},
\end{eqnarray}
and in second order
\begin{eqnarray}
  a_{\bm k}^\dagger a_{-\bm k}^\dagger b_{{\bm k}'}^\dagger b_{{-\bm k}'}^\dagger \ket{0} &=&
  \ket{\bm k}_\text{L} \ket{\bm k}_\text{R} \otimes b_{{\bm k}'}^\dagger b_{{-\bm k}'}^\dagger \ket{0}_{b}, \\
  a_{\bm q}^\dagger a_{-\bm q}^\dagger b_{{\bm k}'}^\dagger b_{{-\bm k}'}^\dagger \ket{0} &=&
  \ket{\bm q}_\text{L} \ket{\bm q}_\text{R} \otimes b_{{\bm k}'}^\dagger b_{{-\bm k}'}^\dagger \ket{0}_{b}, \\
  a_{\bm k}^\dagger a_{-\bm q}^\dagger b_{-\bm k}^\dagger b_{\bm q}^\dagger \ket{0} &=&
  \ket{\bm q}_\text{L} \ket{\bm k}_\text{R} \otimes b_{-\bm k}^\dagger b_{\bm q}^\dagger \ket{0}_{b}, \\
  a_{\bm q}^\dagger a_{-\bm k}^\dagger b_{-\bm q}^\dagger b_{\bm k}^\dagger \ket{0} &=&
  \ket{\bm k}_\text{L} \ket{\bm q}_\text{R} \otimes b_{-\bm q}^\dagger b_{\bm k}^\dagger \ket{0}_{b}.  
\end{eqnarray}
Here, the explicit expression of the tensor product $\ket{0} = \ket{0}_a \otimes \ket{0}_b$ of the individual vacuum
states for photons and intersubband excitations is used and ${\bm k}'$ can be an arbitrary in-plane wave vector.

We carry out the post-selection by projecting onto these states with a projection operator $\mathcal{P}_\text{LR}$,
\begin{equation}
  \ket{\psi_\text{LR}} = \frac{1}{N_\text{LR}} \mathcal{P}_\text{LR} \ket{G^{(2)}},
\end{equation}
with $N_\text{LR}$ being a necessary normalization constant, since the operation is a projection. As an intermediate
result, we obtain the pure state $\ket{\psi_\text{LR}}$ in which all the two-photon states fulfilling the conditions of
the post-selection are extracted. We give an explicit expression for $\ket{\psi_\text{LR}}$ in Appendix A.

As we will see below, the reduced density matrix $\varrho^a$ of the photonic system is needed for the
calculation of the entanglement. We compute $\varrho^a$ by tracing out the intersubband excitations
\begin{eqnarray}
\label{eqn:phot_state}
  \varrho^a &=& \text{Tr}_b \ket{\psi_\text{LR}} \bra{\psi_\text{LR}} \nonumber \\
  &=& \frac{1}{N_\text{LR}^2} \left( \begin{array}{cccc}
               Z(k) & 0 & 0 & Y(k,q) \\
               0 & X(k,q) & 0 & 0 \\
               0 & 0 & X(k,q) & 0 \\
               Y(k,q) & 0 & 0 & Z(q)
             \end{array}
          \right).
\end{eqnarray}
Here, the matrix representation is in the basis (\ref{eqn:basis}) and the abbreviations
\begin{eqnarray}
\label{eqn:abbrev_X}
  X(k,q) &=& |F(k)|^2 |F(q)|^2, \\
\label{eqn:abbrev_Y}
  Y(k,q) &=& G(k) G(q) \times \nonumber \\
  &\phantom{=}& \times \left[ \left( 1 + \frac{1}{2}S \right) - |F(k)|^2 - |F(q)|^2 \right], \\
\label{eqn:abbrev_Z}
  Z(k) &=& X(k,k) + Y(k,k),
\end{eqnarray}
were introduced and $S$ is the sum over all expansion coefficients squared
\begin{equation}
\label{eqn:infinite_sum}
  S = \sum \limits_{{\bm k}'} G^2(k').
\end{equation}
The value of $S$ (see Appendix B) depends on the sample area $A$. In the limit of $A \gg \frac{2\pi}{\mathcal{I}}
\frac{c}{\omega_{12}}$, one can take
the limit $S \rightarrow \infty$ and obtains
\begin{equation}
  \varrho^a \stackrel{S\rightarrow\infty}{=} \frac{1}{G^2(k) + G^2(q)} \left( \begin{array}{cccc}
               G^2(k) & 0 & 0 & G(k) G(q) \\
               0 & 0 & 0 & 0 \\
               0 & 0 & 0 & 0 \\
               G(k) G(q) & 0 & 0 & G^2(q)
             \end{array}
          \right).
\end{equation}
This corresponds to the pure photon state
\begin{equation}
  \ket{\psi^a} = \frac{1}{\sqrt{G^2(k) + G^2(q)}} \left( G(k) \ket{\bm k}_\text{L} \ket{\bm k}_\text{R}
  + G(q) \ket{\bm q}_\text{L} \ket{\bm q}_\text{R} \right).
\end{equation}

\subsection{\label{subsec:quant}Measure of entanglement}

The state $\varrho^a$, which we derived from $\ket{G^{(2)}}$, describes two photons that propagate with opposite
in-plane wave vectors in the microcavity and that can potentially be released by an appropriate time-modulation (quench)
of the Rabi frequency $\Omega_\text{R} (k)$. Since one chooses the modes $\bm k$ and $-\bm k$ and accordingly $\bm q$
and $-\bm q$ via the post-selection, the photons effectively form a two-qubit system. For such a system, the
entanglement for mixed states can be calculated analytically without evaluating a convex roof explicitly from the density
matrix by way of the concurrence $C(\varrho)$ \cite{Wootters98}. With this function, the so-called entanglement of
formation $E_\text{F} (\varrho)$ \cite{Bennett96} of two qubits can be easily calculated via \cite{Wootters98}
\begin{equation}
  E_\text{F} (\varrho) = h \left( \frac{1 + \sqrt{1 - C^2(\varrho)}}{2} \right),
\end{equation}
with the binary entropy
\begin{equation}
  h(x) = - x \log_2 (x) - (1 - x) \log_2 (1 - x).
\end{equation}
The concurrence itself is given by \cite{Wootters98}
\begin{equation}
\label{eqn:conc}
  C(\varrho) = \max \{0, \lambda_1 - \lambda_2 - \lambda_3 - \lambda_4\}.
\end{equation}
Here, $\lambda_1$ to $\lambda_4$ are, in decreasing order, the square roots of the eigenvalues of the matrix $\varrho
\tilde{\varrho}$
and $\tilde{\varrho}$ is a transformation of the density matrix given by \cite{Wootters98}
\begin{equation}
  \tilde{\varrho} = (\sigma_y \otimes \sigma_y) \varrho^* (\sigma_y \otimes \sigma_y),
\end{equation}
where $\sigma_y$ is the Pauli $y$ matrix and the $^*$ denotes complex conjugation.

\subsubsection{Analytical results}
For the photonic state $\varrho^a$ (\ref{eqn:phot_state}), the parameters $\lambda_1$ to $\lambda_4$ are found
to be
\begin{eqnarray}
  \lambda_1 &=&\frac{1}{N_\text{LR}^2} \left( \sqrt{Z(k) Z(q)} + Y(k,q) \right), \\
  \lambda_2 &=& \frac{1}{N_\text{LR}^2} \left( \sqrt{Z(k) Z(q)} - Y(k,q) \right), \\ 
  \lambda_{3,4} &=& \frac{1}{N_\text{LR}^2} X(k,q).
\end{eqnarray}
Hence from (\ref{eqn:conc}), we obtain in connection with (\ref{eqn:abbrev_X})-(\ref{eqn:abbrev_Z}) for the concurrence
\begin{widetext}
\begin{eqnarray}
\label{eqn:conc_result}
 C(\varrho^a) =  C(k,q) = \frac{2}{N_\text{LR}^2} \left( G(k) G(q) \left[ \left( 1 + \frac{1}{2} S \right)
  - |F(k)|^2 - |F(q)|^2\right] - |F(k)|^2 |F(q)|^2\right).
\end{eqnarray}
\end{widetext}
The concurrence thus only depends on the absolute values $k$ and $q$ via the expansion coefficients, which were given in
(\ref{eqn:exp_coeff}) and (\ref{eqn:exp_coeff_F}),
\begin{eqnarray*}
  G(k) &=& \frac{\omega_{12} + \omega_\text{c}(k) - \omega_\text{LP}(k) - \omega_\text{UP}(k)}{\omega_{12} -
  \omega_\text{c}(k) - \omega_\text{LP}(k) - \omega_\text{UP}(k)}, \\
  |F(k)| &=& \frac{\omega_{12}}{\Omega_\text{R} (k)} G(k).
\end{eqnarray*}

To show the dependance of the concurrence on $k$ and $q$, we have to evaluate the sum $S$ explicitly. Using a sample
area $A = (200\,\mu\text{m})^2$, we find $S = 0.716$.
\begin{figure}
  \includegraphics[width=\columnwidth]{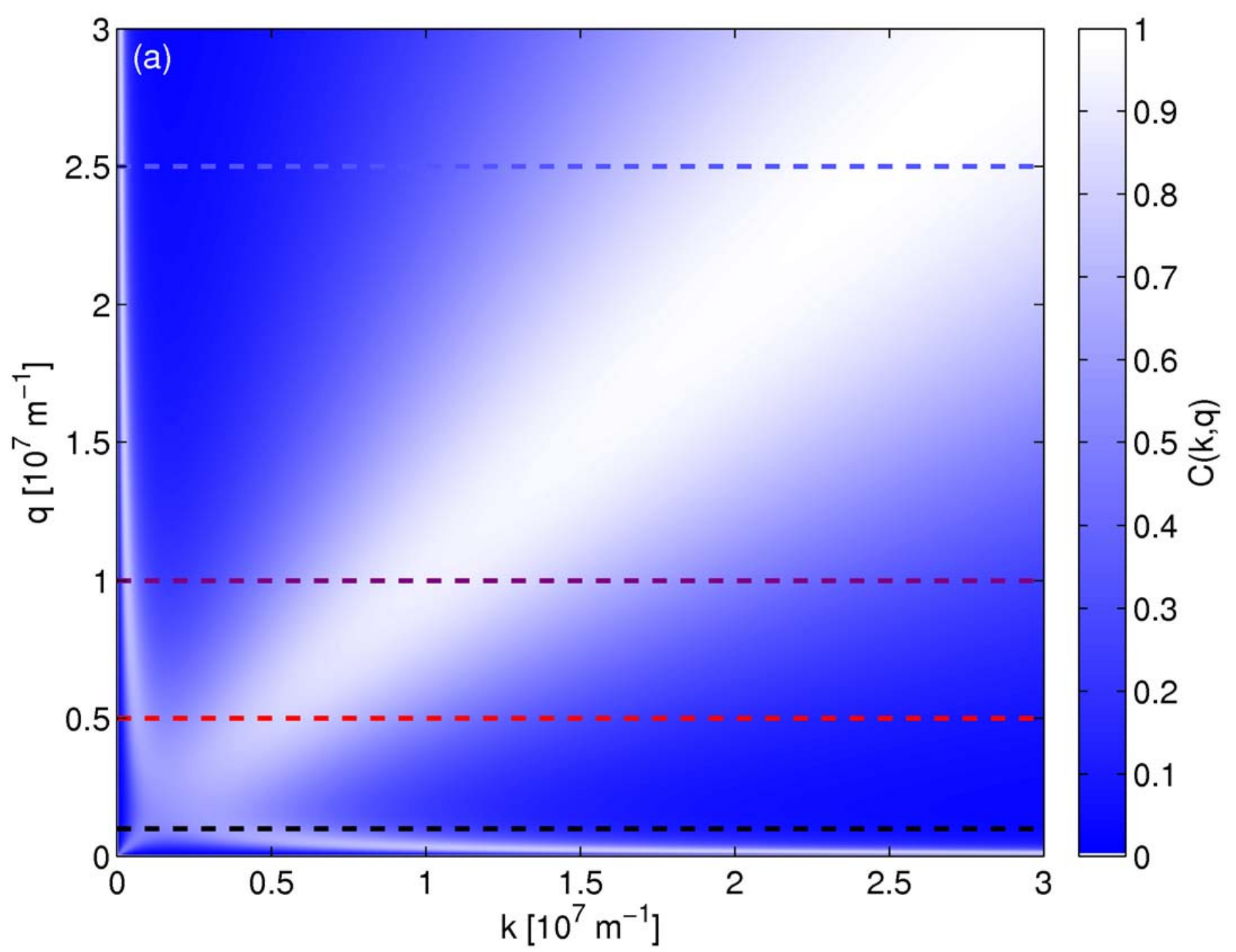}
  \includegraphics[width=\columnwidth]{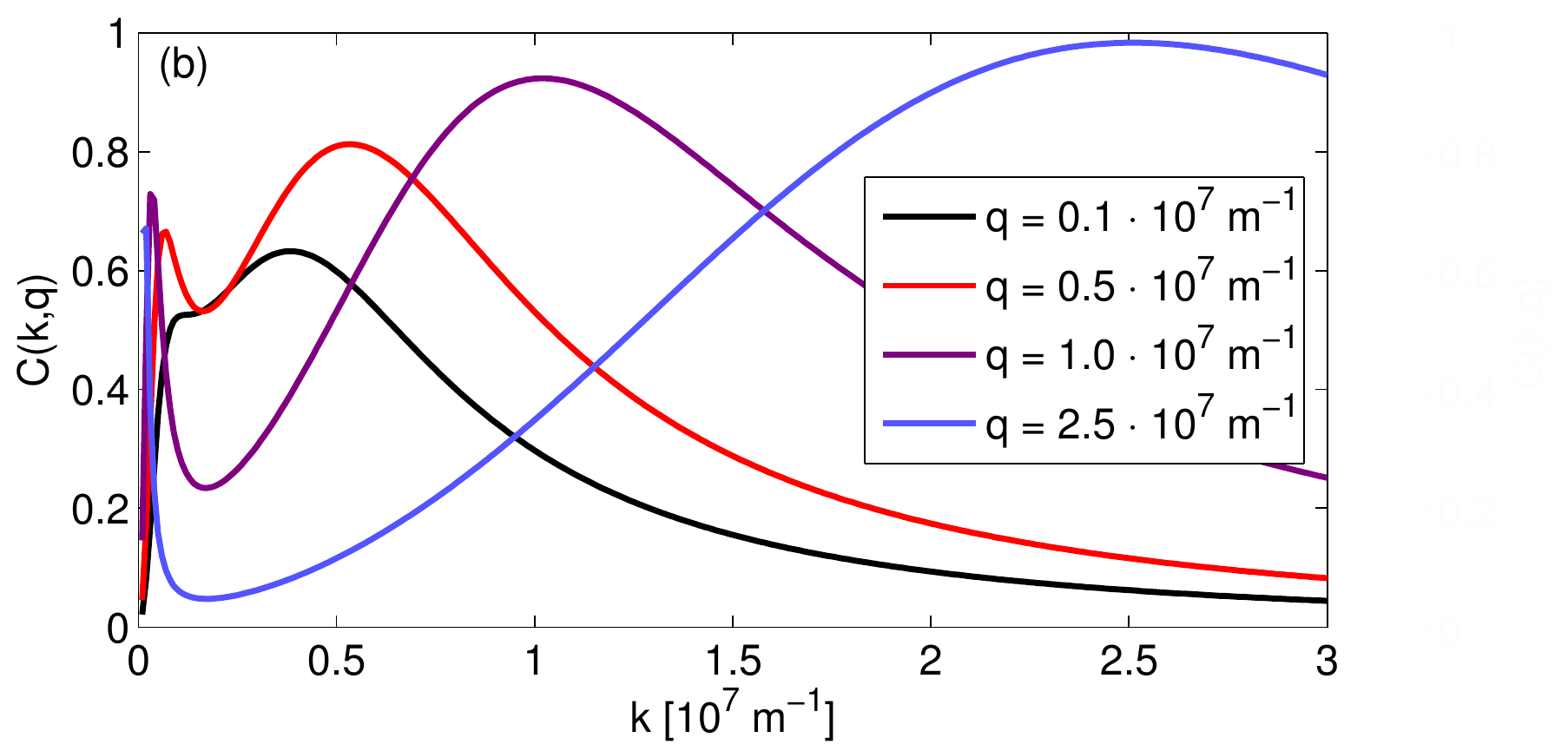}
  \includegraphics[width=\columnwidth]{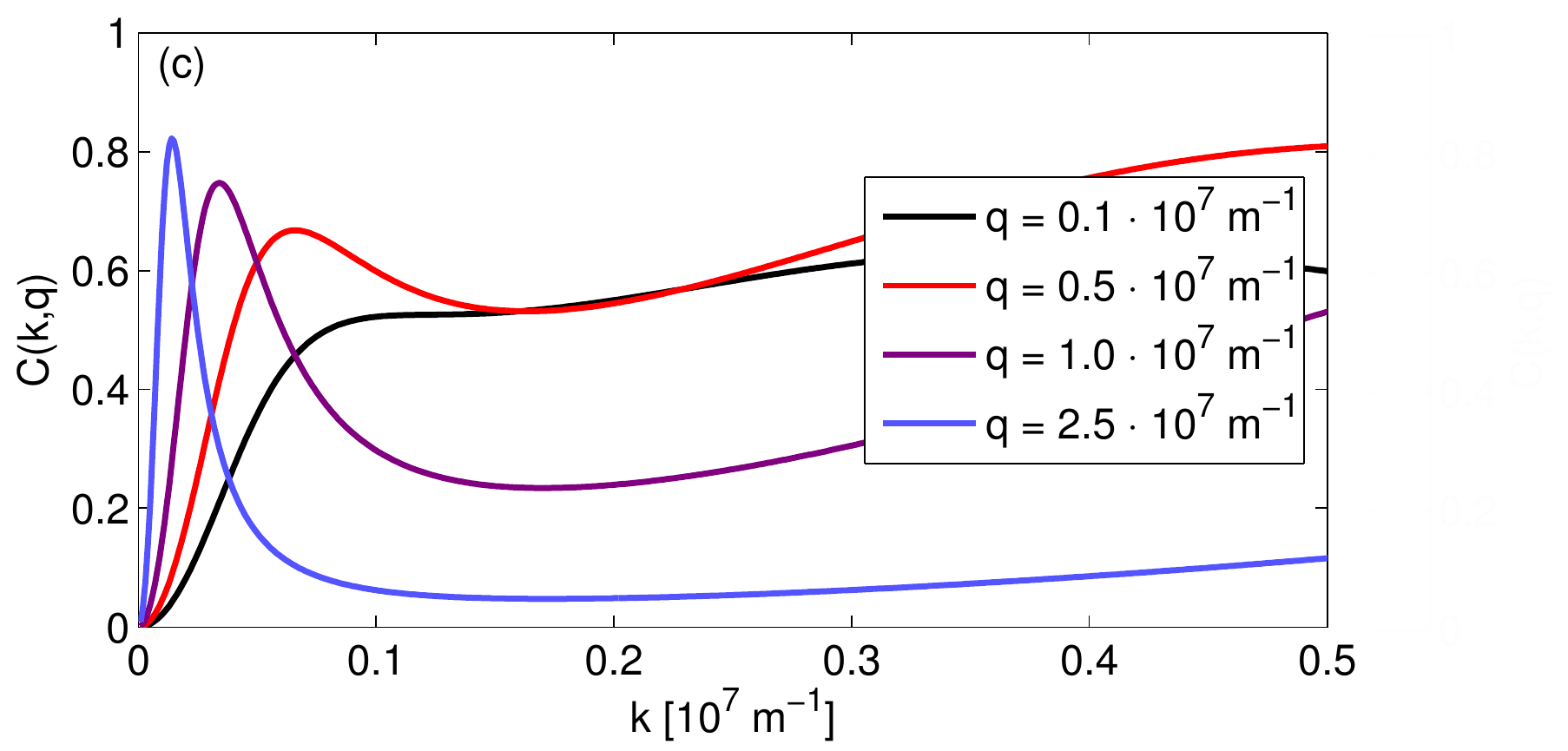}
  \caption{\label{fig:conc}(a) The concurrence $C(k,q)$ as a function of $k$ and $q$ for GaAs/AlGaAs quantum wells.
  (Parameter values: $\varepsilon = 10$, $f_{12} = 14.9$,
  $n_\text{QW}^\text{eff} = 50$, $L_\text{c}^\text{eff} = 2\,\mu\text{m}$, $\hbar \omega_{12} = 150\,\text{meV}$,
  $N_\text{2DEG} = 10^{12}\,\text{cm}^{-2}$.) In (b), we plot $C(k,q)$ for fixed values of $q$ (given in the respective
  legend) and for a large range of $k$ from 0 to $3 \cdot 10^{7}\,\text{m}^{-1}$, which corresponds to a photon energy
  of 1.9$\,$eV (450$\,$THz, red) and in (c) magnification of the range up to $k = 0.5\cdot 10^{7}\,\text{m}^{-1}$
  (320$\,$meV, 80$\,$THz, mid infrared).}
\end{figure}
The result is presented in Fig.~\ref{fig:conc} for the same parameters as used before, where $C(k,q)$ is shown
in a density plot as a function of the modes $k$ and $q$. Below it, we show cuts for
different values of $q$ to illustrate the dependancy of the concurrence better. 
One can observe two branches of high entanglement that
appear for large values of $k$ and/or $q$. Their appearence can be explained by the characteristics of the
expansion coefficient $G(k)$ for large $k$. $G(k)$ tends to zero as $1/k^2$, hence $|F(k)|^2$ scales as $1/k^3$.
Hence for the diagonal branch, i.e. $k \approx q$, the $|F|$ terms in (\ref{eqn:conc_result}) and (\ref{eqn:norm}) can
be neglected and we obtain
\begin{equation}
\label{eqn:conc_large_k}
  C(k,q) \stackrel{k,q \rightarrow \infty}{\approx} \frac{2 G(k) G(q)}{G^2(k) + G^2(q)} \stackrel{k \approx q}{\approx}
  1.
\end{equation}
Accordingly, photons in the visible regime are almost maximally entangled if their wave numbers are of the same size.

The other branch appears if $G(k) \approx G(q)$ and the modes are far from each other. We give a more precise analysis
of expression (\ref{eqn:conc_large_k}) in the next subsection, where the limit of large sample areas is worked out.

\subsubsection{Large-cavity limit}
\begin{figure}
  \includegraphics[width=\columnwidth]{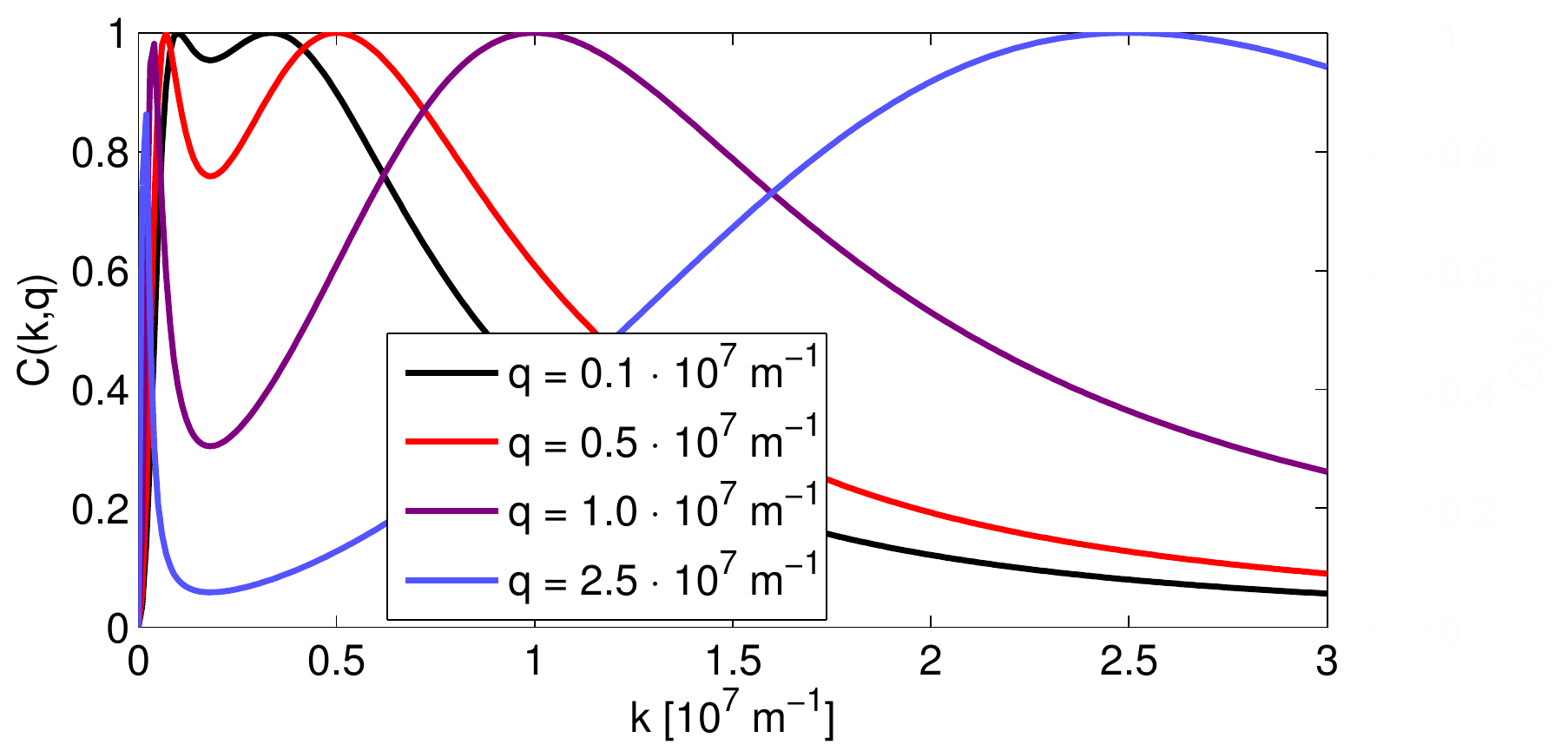}
  \includegraphics[width=\columnwidth]{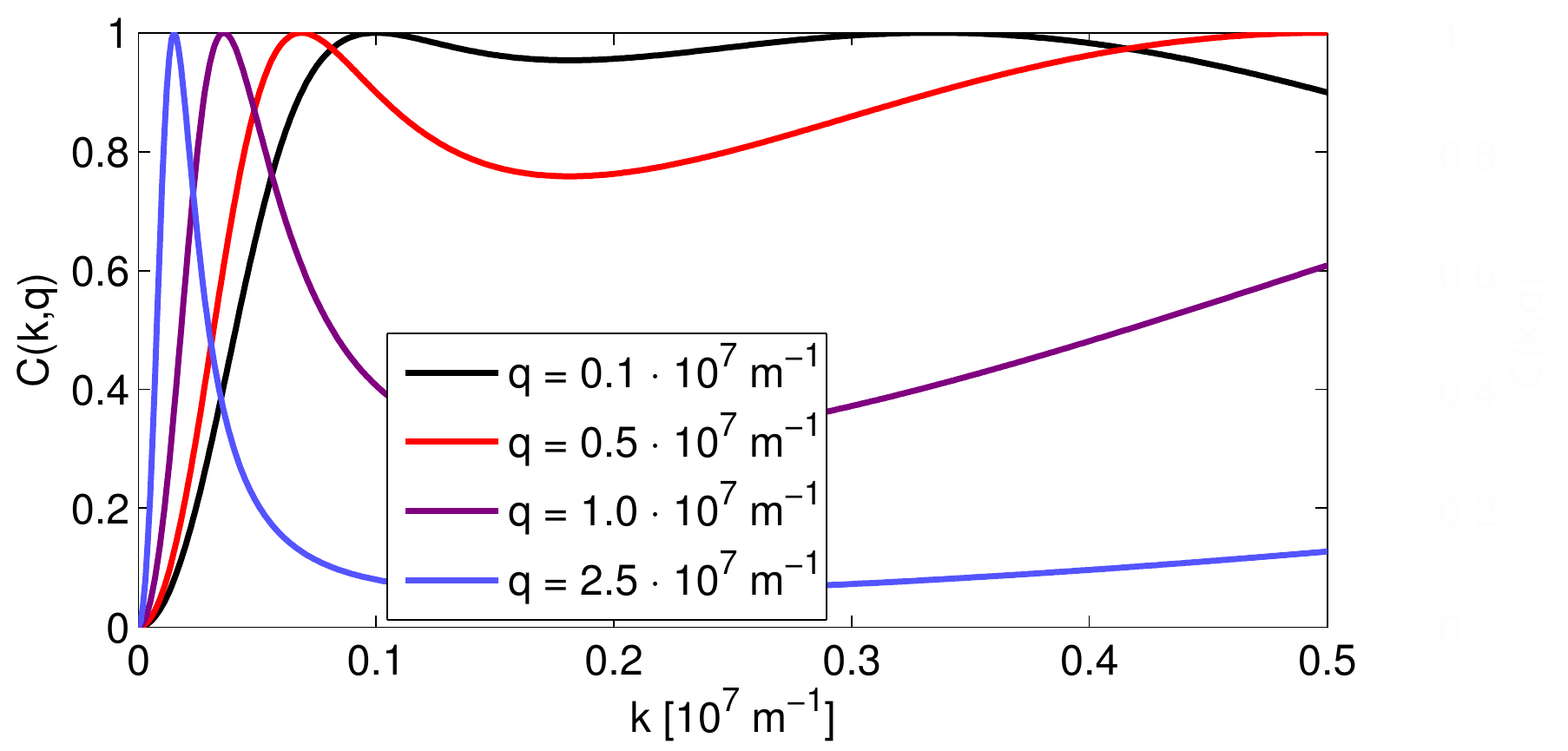}
  \caption{\label{fig:conc_limit}The concurrence $C(k,q)$ in the case of large sample areas for GaAs/AlGaAs quantum
  wells.
  (Parameter values: $\varepsilon = 10$, $f_{12} = 14.9$,
  $n_\text{QW}^\text{eff} = 50$, $L_\text{c}^\text{eff} = 2\,\mu\text{m}$, $\hbar \omega_{12} = 150\,\text{meV}$,
  $N_\text{2DEG} = 10^{12}\,\text{cm}^{-2}$.)}
\end{figure}

The case of a large cavity, i.e. a large sample area $A$, is described by the limit $S \rightarrow \infty$, see
Appendix B. The concurrence in this case is calculated to be
\begin{equation}
  C(k,q) = \frac{2 G(k) G(q)}{G^2(k) + G^2(q)}.
\end{equation}
We show the result in Fig.~\ref{fig:conc_limit}. As one can see, the
concurrence always has two maxima if $q$ is held constant. One maximum appears for $G(k) = G(q)$ and $k
\neq q$. There, photons are maximally entangled since we have $C = 1$. However, this maximum is relatively
sharpely peaked and if one realizes the post-selection experimentally by choosing a certain finite $k$-range,
the entanglement will be reduced. The other maximum appears when $k =
q$, which seems to be an artefact of the calculation, since this case was excluded in the calculations above.
The reason for the exclusion is that the two-photon states would be seperable and hence not entangled. In
this case, however, the maximum is quite broad. Therefore, for a given $q$ there exists a wide range of
corresponding modes $k$, for which the two photons are almost maximally entangled.
In the second plot of Fig.~\ref{fig:conc_limit} we show
a magnification for wave vectors up to $0.5 \cdot 10^{7}\,\text{m}^{-1}$ to show the dependance of
$C(k,q)$ on $k$ clearer around the first maximum, which is not visible in the previous plot. In particular
at the intersubband resonance, which is around $0.2 \cdot 10^{7}\,\text{m}^{-1}$, the two maxima approach each other
so that by selecting different modes around the resonance, the entanglement of the photons can be made
almost maximal. The corresponding photon energies are in the mid infrared, about 150$\,$meV.

\section{\label{sec}Conclusion}

An efficient and deterministic source of entangled photons is needed in quantum information processing. In
this work, we examined a new scheme of photon production, based on the emission of quantum vacuum radiation
from the intersubband cavity system. Because the triggered photon emission is based on a non-adiabatic
modulation of the system's ground state, an exact expression for this state could be used. Since the ground state
consists of an infinite number of photonic and electronic states, we propose a post-selective measurement to
reduce the photonic system to an effective two-qubit system, in which the qubit state was defined as two
different in-plane wave vectors. The so-called mode entanglement of the photons is quantified by the
concurrence. We found an analytical expression for the concurrence, which depends on the absolute values of the
chosen wave vectors. We find that the concurrence, and therefore the entanglement of the post-selected photons, is
non-zero. In the
limiting case of large sample areas, there exists a continuous set of mode pairs for which the concurrence is 1,
i.e.~the photons are maximally entangled. Also, in this case it turns out that for photon energies
around the intersubband resonance, which is in the mid infrared regime of the electromagnetic spectrum,
the photons are almost maximally entangled, the concurrence being close to 1. This is fundamentally
important for the possible use in quantum information processing. Furthermore, a high degree of entanglement
can be achieved if the modes chosen in the post-selection are close to each other. Therefore, one could extract
entangled photon pairs in technologically relevant frequency domains like a telecom wavelength.
\begin{acknowledgments}
  The authors thank A. Pashkin and R. Huber for useful discussion and acknowledge funding from the DFG within SFB 767.
\end{acknowledgments}

\appendix

\section{Post-selected state}
The pure state $ \ket{\psi_\text{LR}}$, which contains all two-photon states fulfilling the conditions of the
post-selection, is given as
\begin{widetext}
\begin{eqnarray}
  \ket{\psi_\text{LR}} &=& \frac{1}{N_\text{LR}} \mathcal{P}_\text{LR} \ket{G^{(2)}} \nonumber \\
  &\phantom{=}& \frac{1}{N_\text{LR}} \Bigg[ \left( G(k) \ket{\bm k}_\text{L} \ket{\bm k}_\text{R} +
  G(q) \ket{\bm q}_\text{L} \ket{\bm q}_\text{R} \right) \otimes
  \left( \ket{0}_b + \frac{1}{2} \sum \limits_{{\bm k}'} G(k') b_{{\bm k}'}^\dagger b_{{-\bm k}'}^\dagger
  \ket{0}_{b} \right) + F^2(k) \ket{\bm k}_\text{L} \ket{\bm k}_\text{R} \otimes b_{-{\bm k}}^\dagger b_{{\bm
  k}}^\dagger \ket{0}_{b} \Bigg. \nonumber \\
  &\phantom{=}&\phantom{\frac{1}{N_\text{LR}}}
  + F^2(q) \ket{\bm q}_\text{L} \ket{\bm q}_\text{R} \otimes b_{-{\bm q}}^\dagger b_{{\bm
  q}}^\dagger \ket{0}_{b} + F(k) F(q) \ket{\bm q}_\text{L} \ket{\bm k}_\text{R} \otimes b_{-{\bm k}}^\dagger b_{{\bm
  q}}^\dagger \ket{0}_{b} + F(k) F(q) \ket{\bm k}_\text{L} \ket{\bm q}_\text{R} \otimes b_{-{\bm q}}^\dagger b_{{\bm
  k}}^\dagger \ket{0}_{b} \Bigg. \Bigg]
\end{eqnarray}
with $N_\text{LR}$ being a normalization constant,
\begin{equation}
  \label{eqn:norm}
  N_\text{LR}^2 = \bra{G^{(2)}} \mathcal{P}_\text{LR} \ket{G^{(2)}} = \left( G^2(k) + G^2(q) \right) \left( 1 +
\frac{1}{2} S \right)
  + \left(|F(k)|^2 + |F(q)|^2\right)^2 - 2 G^2(k) |F(k)|^2 - 2 G^2(q) |F(q)|^2.
\end{equation}
\end{widetext}
We already introduced the sum $S$ as being
\begin{equation}
  S = \sum \limits_{{\bm k}'} G^2(k').
\end{equation}

\section{Continuum limit}

When taking the sum over all two-dimensional in-plane wave vectors $\bm k$ in Eqn.~(\ref{eqn:infinite_sum}),
the appearing vectors depend on the boundary conditions. We choose peridodic boundary conditions, and hence
\begin{eqnarray}
  k_x &=& \frac{2\pi}{L_x}n_x,\quad n_x = 0, \pm 1,\dots, \\
  k_y &=& \frac{2\pi}{L_y}n_y,\quad n_y = 0, \pm 1,\dots,
\end{eqnarray}
where $L_{x(y)}$ is the cavity length in $x(y)$-direction and $n_{x(y)}$ is an integer. Every discrete wave
vector $\bm k$ has a volume $\Delta$ in $\bm k$-space:
\begin{equation}
  \Delta = \Delta k_x \Delta k_y = \frac{(2\pi)^2}{L_x L_y} = \frac{(2\pi)^2}{A}
\end{equation}
and $\Delta k_{x(y)}$ is the difference between two adjacent wave vectors in $x(y)$-direction, $A$ the sample area. In
the continuum
limit, the $\bm k$ vectors lie close in the reciprocal space and the sum can be replaced by an integral
\begin{eqnarray}
  S =\sum \limits_{\bm k} G^2(k) &=& \frac{1}{\Delta} \sum \limits_{\bm k} \Delta G^2(k)
  \rightarrow \frac{1}{\Delta} \int \text{d}^2 k \, G^2(k) \nonumber \\
  &=& \frac{A}{2\pi} \int \limits_{k = 0}^{\infty} \text{d} k \, k \, G^2(k) \nonumber \\
  &=& \frac{A}{2\pi}  \left( \frac{c}{\omega_{12}} \right)^2 \mathcal{I},
\end{eqnarray}
where we use polar coordinates to evaluate the integral and carried out the polar-angle integration. In the last step,
we make a substitution and introduce the dimensionless variable $\tilde{k} :=
\frac{c}{\omega_{12}} k$ to get the dimensionless integral $\mathcal{I}$,
\begin{equation}
  \mathcal{I} = \int
\limits_{\tilde{k} = 0}^{\infty} \text{d} \tilde{k} \, \tilde{k} \, G^2\left(\frac{c}{\omega_{12}}\tilde{k}\right).
\end{equation}
One can show that the expansion coefficient $G(k)$ decreases like $1/k^2$ for large $k$. Consequently, the integrand has
the asymptotics
\begin{equation}
  k \, G^2(k) \stackrel{k \rightarrow \infty}{\approx} \frac{1}{k^3}
\end{equation}
and hence, the integral converges. We evaluate $\mathcal{I}$ numerically using the same parameters as above and find
$\mathcal{I} = 1.9 \cdot 10^{-4}$.

\end{document}